# Quantitative Approach to Intensity-Modulated Radiation Therapy Quality Assurance Based on Film Dosimetry and Optimization

Dong Hyun Park[a)], Sung-Yong Park [b)], Dahl Park, Tae-Hyun Kim, Kyung Hwan Shin, Dae Yong Kim and Kwan-Ho Cho

Center for Proton Therapy, National Cancer Center, Goyang 411-769, Korea

## Abstract

To accurately verify the dose of intensity-modulated radiation therapy (IMRT), we have used a global optimization method to investigate a new dose-verification algorithm. In practical application of this quality assurance (QA) procedure, verification of the dose using calculated and measured dose distributions involves a subtle problem in the region of high dose gradient. Consideration of systematic errors shows that the large dose differences in high-dose-gradient regions are due to the unexpected shift of measuring devices. We have proposed an optimization algorithm to correct this error, and an optimization method to minimize the average dose difference has been used in this study. The relationship between the dose-verification procedure and the applied optimization algorithm is explained precisely. Optimization dramatically reduced the difference between measured and calculated dose distributions in all cases investigated. The obtained results support the relevance of our explanations for the problem in the high-dose-gradient region. We have described this dose-verification procedure for IMRT and intensity-modulated radiosurgery. Through this study we have also developed an intuitive reporting method that is statistically reasonable.

## 1. INTRODUCTION

The accuracy and capabilities of technologies for radiation therapy planning and treatment delivery have recently improved dramatically, and this improvement has naturally led to increasingly complex patient-specific quality assurance (QA) procedures. Intensity-modulated radiation therapy (IMRT) is one of the main applications of the complicated patient specific QA [1]. IMRT is usually delivered by complicated motions of a multileaf collimator (MLC) that generates patient-specific nonuniform intensity distributions. The patient specificity and highly sophisticated nature of the beam profiles lead to a requirement for pretreatment QA of IMRT. The well-known patient-specific IMRT QA consists of two main steps: (1) the measurement of absolute point doses using an ionization chamber, and (2) 2D relative dosimetry using a film, beam imaging system [2], electronic portal imaging device (EPID) [3–5], or another technique. These IMRT QA procedures are very time consuming, labor intensive, and hence expensive, which limits the number of patients on which they can be used. The development of a computerized QA procedure represents a promising solution to the aforementioned problems, since it should reduce the implementation time and cost without loss of accuracy. Existing computerized QA tools such as an independent dose-calculation algorithm [6–10], a verification of the distribution using several 2D dosimeters, and an algorithm for direct checking of the MLC fluence [11], validate the usefulness of the computer-based approach.

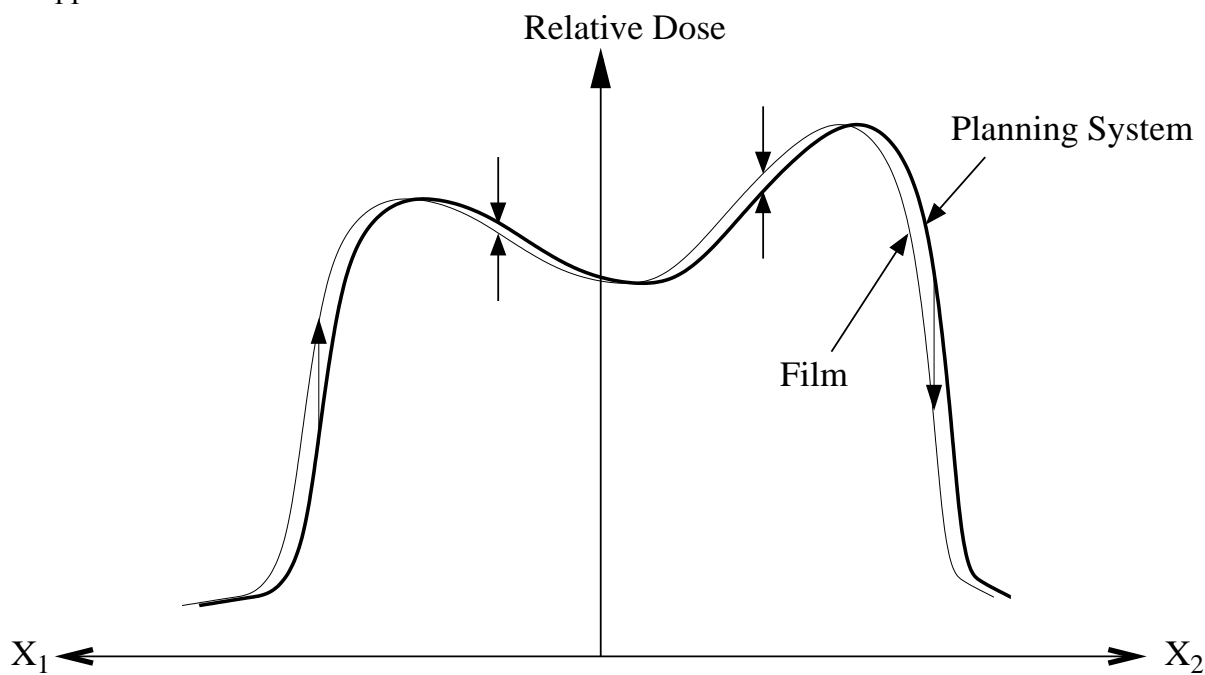

Fig. 1. Schematic diagram showing that a small shift of the dosimeter generates larger dose differences at regions with a higher dose gradient.

When physicists perform patient-specific QA procedures, especially the dose-verification procedure, they find that the regions with a high dose gradient are very difficult to handle. Such regions in the IMRT field are highly sensitive to movement of the dosimeters, and thus small setup errors due to the tolerance of the positioning device can represent a serious problem. Positioning devices, especially lasers, pass to the isocenter of a linear accelerator (LINAC) vertically, axially, and laterally. In practical applications the isocenter of a LINAC does not coincide with a point, but is floated within the LINAC tolerance. The laser indicates the isocenter with the finite size width, thus the tolerance of laser cannot be separated from the LINAC tolerance. The small translations and rotations within the tolerance of the LINAC are the only possible mechanical degrees of errors when we position the phantom. From a 2D dosimetric viewpoint the film is localized in the isocenter surface, and therefore significant setup errors come from translations parallel to the film. The effect of the small setup errors – particularly a tiny shift – in the regions with a high dose gradient is illustrated in Fig. 1. One purpose of 2D dosimetry is to verify the treatment delivery as simulated in the planning system. The small shift within the tolerance of LINAC means that this problem is unavoidable within conventional schemes for verifying the dose distribution, which is why we can encounter unexpected results near high-dose-gradient regions in the dose-verification procedure.

Dose verification using the γ-index method [12–14] is well known to resolve this problem. The γ-index method has suggested dual criteria for the low (e.g., 3% dose differences) and the high (e.g., 3-mm distance) dose-gradient regions. The γ-index is formulated such that when it is smaller than one, either the dose difference or the distance is less than its criterion, and when it is larger than one, either the dose difference or the distance is larger than its criterion. In other words the patient plan is accepted when $\gamma \leq 1$ and rejected when $\gamma > 1$. The result of dose verification using this method includes all the systematic and random errors that are generated by the measurement procedure.

The other aggressive way to solve the problem is to precisely investigate the reason for the unexpected shift, and then to recover it to the correct position. Optimization involves finding the solution to a problem with constraints, and fortunately this method can also be applied to our problem. For the verification of an actual patient-specific dose distribution, it is obvious that a larger setup error induces a larger average dose difference (Fig. 1). This means that the 2D function generated by averaging the absolute dose differences of each of all possible 2D translations is always bowl shaped and concave up, and hence has a global minimum. This 2D function is called an objective function, the average of the absolute dose differences is an optimization parameter, and the process for finding the minimum is the optimization path of the problem. Generally the optimization algorithms are chosen according to their efficiency for a specific purpose, and thus developing a useful optimization algorithm (i.e., finding the effective optimization path) is the most important task in solving the problem. After the optimization, the solution shows us the average of the absolute dose differences of the global minimum and the distance between the optimized position and the installed phantom. Here the distance is merely the unexpected shift of the dosimeter which has been screened by the LINAC tolerance. Verifying the dose distribution becomes meaningful when this distance is smaller than the tolerance of the LINAC.

In fact this method does not conflict with the γ-index method. Since the purpose of optimization is simply subtraction of the systematic error due to the tolerance of the positioning device or LINAC, we can apply the γ-index method after the optimization procedure. In this article we introduce a new dose-verification scheme that appears faster than other methods and is also clinically useful. Statistical reliability and an intuitive interface are also strong points of this scheme. The specific representation of optimization algorithms is discussed here rigorously in terms of patient treatment methods such as IMRT and intensity-modulated radiosurgery (IMRS).

## 2. MATERIALS AND METHODS

Of the many 2D dosimeters available, we concentrate on film dosimetry in this study since it is one of the most convenient ways of performing 2D dosimetry. The setup of a film and solid water phantom are easier than the other methods, and convenience of maintenance and low cost are also attractive points. On the other hand the narrow air cavities inside film jacket can generate an unexpected build-up behavior of the beam. Overresponse for low-energy photons is a significant problem in film dosimetry, but this symptom can be resolved by attaching thin lead filters [15]. Calibration for optical density (OD) and distortion depends entirely on how many films are used and how frequently it is performed. All the above factors can affect the accuracy of dose-distribution verification; among the possible candidates, Kodak Extended Dose Rate (EDR2) films were chosen since they were considered to minimize the above problems.

Details of the QA procedure at our institution are as follows. The first step is to obtain the MLC sequencing files from the patient treatment plan, and to apply these files to a water phantom inside the planning system. This simulation provides us with the specific dose-distribution files for the water phantom in the planning computer. The next step is to measure the absolute dose using a pinpoint ion chamber, and to perform the 2D film dosimetry using EDR2 films. The solid water phantom (PLASTIC WATER, Nuclear Associate) is positioned as in the simulation, which is called the hybrid method [16, 17]. The ion chamber and film are located at a depth of 5 cm in the solid water phantom, and its source-to-surface distance (SSD) is 95 cm. We note that the vertically and axially aligned films for obtaining the dose distribution (so-called star-shot) are irrelevant in the cases of patient QA with only coplanar beams, since the film dominantly contains the MLC contribution for isocenter slice leafs instead of entire leafs of fields. Occasionally the position of the pinpoint chamber is chosen at a low-dose-gradient region to avoid the aforementioned problems associated with high-dose-gradient regions. The obtained dose using the ion chamber should agree with that of the planning system within ±3%. This measurement of the absolute point dose can be replaced by the independent monitor unit calculation program in order to reduce the QA implementation time. When we install the film, its isocenter is positioned using a positioning laser, and the isocenter is punctured or marked by a needle or other device. Individual IMRT fields are subsequently delivered to the ion chamber and film, respectively, and this provides us with the absolute dose of a specific point and the delivered dose-distribution file after developing the film. As described above, the laser is associated with an intrinsic uncertainty due to the tolerance of the LINAC isocenter, which is typically ±2 mm for IMRT devices and ±1 mm for radiosurgery devices. This indicates that film dosimetry has an unavoidable uncertainty even if the mechanical tolerance is reduced using other correction methods.

The remaining procedures are related to dose verification, and these are performed by the verification computer. We have used homemade software modules for managing dose-distribution files, optimizing the position, calculating dose differences, and printing out the report.

The first step of the computerized procedure is to obtain useful dose distributions from the raw dose-distribution files. The dose-distribution file calculated using the CadPlan planning system (Varian, R.6.3.6) contains the point doses at spacings of 1.25, 2.5, or 5 mm in both *X* and *Y* directions, with the resolution obviously depending on the planning system used. Linear interpolation of the given point doses produces a more precise dose distribution at a 1-mm resolution in both directions. For the dose distribution measured using film, we obtain the inline multiple profiles with 5-mm spacing using the Omni-Pro (Wellhöffer, Ver.6.0A, Germany) software module with a scanner (VIDAR, VXR16). After obtaining the profiles, point doses at 5-mm intervals are taken for each profile, producing the measured dose distribution with a 5-mm resolution. There are several ways of determining the resolution of the measured dose distribution, depending on how many points we want to verify. For example, the 5-mm resolution of the measured dose distribution gives us over 400 points in the case of a field size of 10 x 10 $cm^2$. Although we are able to obtain a smoother distribution, several hundred verification points provide sufficient data for the statistical analyses and are sufficiently few for the calculation to be completed within an acceptable time frame. The number of verification points is generally determined by the required implementation time.

The second step is to apply the optimization algorithm to recovering the unexpected shift of the dosimeter. A usual way of verifying the dose distribution involves steps for overlapping the measured and calculated dose distributions, and comparing point doses at all matched positions. Therefore, obtaining measured and calculated dose distributions at the same resolution seems indispensable. The reason to obtain dose distributions with different resolutions is to apply the optimization procedure. Let us consider an example optimization algorithm. First, the measured and calculated dose distributions are superimposed by coincidence of their origins, after which the measured dose distribution is translated to a randomly selected position on the calculated dose distributions. Every point dose in the translated distribution that we want to verify is compared with the corresponding point doses of the calculated dose distribution. This step provides us with the average of the absolute dose difference corresponding to that translation. Calculations of the average of the absolute dose difference near the already translated positions are used to determine the most promising direction toward the optimized position, and this will generally be in the direction of a smaller difference value. From this information we determine the direction of the next translation. The application of several translation-and-comparison iterations produces the global minimum associated with the minimized average dose difference. This is just one example of the optimization algorithm. The distance between the global minimum and the origin of the calculated dose distribution corresponds to the unexpected shift of the 2D dosimeter that we want to recover, and it should be smaller than the tolerance of the LINAC in each direction. This explains why we use a resolution of 1 mm for the calculated dose distribution – this resolution is the minimum translation unit in the optimization algorithm. If this resolution were larger than the LINAC tolerance, then the argument that we verify the calculated dose distribution within the tolerance of the LINAC is irrelevant even if the global minimum is found in the minimal translation. Smaller resolutions are therefore highly recommended for finding the global minimum position more accurately, but of course this is restricted by the available computing resources.

IMRS treatments use very small size of fields, and the dose gradients in penumbrae are much higher than normal IMRT fields. IMRS therefore requires more careful treatment delivery than IMRT, and a tolerance of ±1 mm is typically used. In the optimized dose-verification procedure, the resolution of the measure dose distribution is still taken to be 5 mm, and that of the calculated one is restricted to at most 0.5 mm in order to obtain a result within the LINAC tolerance. A survey of optimization procedures for IMRT and IMRS reveals that both treatments have the same resolution for the measured dose distribution and a different resolution for the calculated one. These features are due to the calculated dose distribution determining the accuracy of the optimization algorithm, and that of the measured distribution deciding the statistical accuracy of the dose-verification procedure.

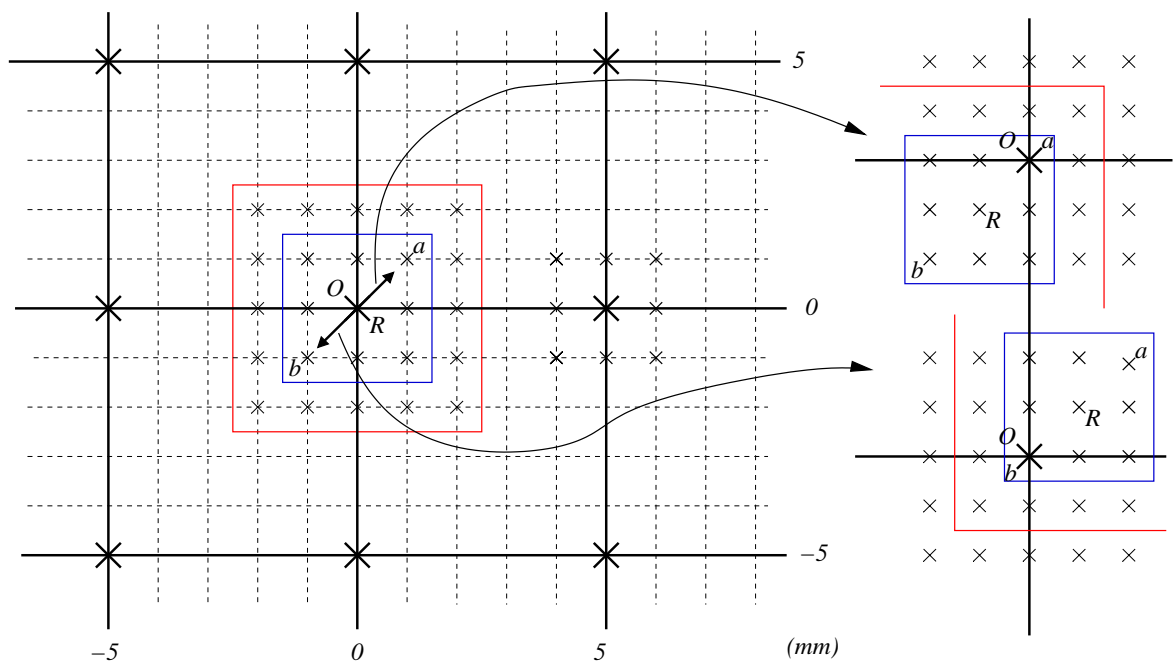

Fig. 2. Schematic diagrams of the shift of measured dose distributions. The overlap between the measured and calculated dose distributions are illustrated in the left panel. Large **X**'s in the bold grid denote the dose distribution measured at a 5-mm resolution, and small x's in the dashed grid denotes the distribution calculated at a 1-mm resolution. Here *O* and *R* are the isocenters of the measured and calculated dose distributions, respectively. The translation of the film to points *a* and *b* is shown in the right panel. The small x points inside the square boxes indicate nearest-neighbor (inner square) and next-nearest-neighbor (outer square) points. Comparison of all large **X** points with matched small x points gives us the average of the absolute dose difference of the corresponding film translation.

There are many possible optimization algorithms, each of which is chosen on the basis of its efficiency and accuracy in a particular application. In our case the shift of the film is expected to be less than the LINAC tolerance, which makes it possible to greatly simplify the aforementioned optimization algorithm. Obviously, comparing all possible translations will always produce the global minimum, but generally this is extremely inefficient. Because it is believed that the optimized position for the film is near the measured position marked on the film, the comparison of the average of absolute dose differences obtained from sufficient translations near the measured position provides us with the global minimum (Fig. 2). Contrary to usual cases, these limited translations are very effective in our case.

After finding the optimal translation, we should rescale the measured dose distribution using the previously measured absolute point dose and the prescription dose. The obtained absolute point dose at a specific position is already verified from the calculated result. A coincidence of doses at a specific point of the measured dose distribution and the ion chamber result provides us with the absolute measured dose distribution. Generally this coincidence also gives us the relative dose with respect to the prescription dose, but assuming this without taking account of other considerations can cause a serious problem, as follows. The isocenter of IMRT or IMRS cases is usually localized in the center of mass of the target, so that dosing the isocenter with 100% of the prescription dose makes sense. This naturally leads to normalization of the measured dose distribution to a value of 100% at the isocenter. However, in some cases of head and neck cancer, the planning system can assign the isocenter to near the spinal cord, which means that we cannot normalize the isocenter dose to the 100% value. There are many ways to determine the isocenter dose, but the most reliable way is to take it directly from the patient plan. The dose differences could be significantly magnified if the aforementioned rescaling is not considered, which would make the dose-verification result irrelevant.

We now discuss the weighting factor introduced to define the clinical weight applied to a specific point. When comparing dose differences we allow for the absolute dose; for example, a 3% dose difference in a 6-Gy dose region is considered more important than a 3% dose difference in a 2-Gy dose region. This means that if the dose of some reference point is normalized to 100% of its prescription dose, then points receiving 100% and 50% doses are assigned weights of 1 and 0.5, respectively. In other words, a 2% difference in a 100% dose is considered the same as a 4% difference in a 50% dose. This is also meaningful from an absolute dosimetric viewpoint, because 6 Gy with a 2% difference is the same as 3 Gy with a 4% difference in absolute value. The weighted dose difference at the point $X = i$, $Y=j$ is defined as

$$\Delta_{ij} = \frac{D_{ij}^{f}}{100}\left(D_{ij}^{f} - D_{ij}^{p}\right), \tag{1}$$

where $D_{ij}^{f}$ and $D_{ij}^{p}$ are the doses of the measured and calculated dose distribution at the point $X=i$,

$Y=j$. Therefore we can obtain the average of the weighted absolute dose difference as

$$\widetilde{\Delta}_w = \frac{1}{N} \sum_{ij} |\Delta_{ij}|, \qquad (2)$$

where $N$ is the total number of verified points. The average of the weighted absolute dose differences $\widetilde{\Delta}_w$ is used as the optimization parameter, and should be included in the patient report. Here we note that $\widetilde{\Delta}_w$ is not the unique optimization parameter in a given objective; the standard deviation can also be the optimization parameter, because it is believed to provide an objective function that has the same global minimum. Any parameter that provides us a similar objective function can be considered as a candidate for the optimization parameter.

The dose-verification method discussed above produces a small average value, but this value is strongly correlated with the number of verified points $N$. Since the weighted dose differences are negligible in the low-dose region, a larger consideration of this region produces a smaller average value. This means that the weighted average value does not represent a meaningful independent parameter. The average value must be used with another parameter that can control the number of points or the dose range. We therefore introduce the low-dose cutoff (LC) and ignore doses lower than this. This is consistent with low-dose regions of the dose distribution being unimportant clinically. It is obvious that the weighted (or not) average values with different LC values will not be comparable with each other, and hence we cannot determine acceptable criteria for the dose-verification procedure without determining LC.

## 3. RESULTS

In the case of IMRT, we used the Clinac 21EX (Varian Oncology System) with 120 MILLENNIUM MLC and CadPlan with Helios inverse planning modules for the treatment and planning of patients. IMRS employed an M3 system (BrainLab, Germany) with a Clinac 600C/D (Varian Oncology System) and BrainScan (BrainLab) planning system. As mentioned above, EDR2 film and Omni-Pro software module with a VIDAR scanner were used to analyze the film. The film was located at a depth of 5 cm in the solid water phantom, and the SSD was 95 cm. The film calibration for OD was performed at 60, 180, 300, 420, and 540 cGy with 6- and 15-MV photon beams. We applied our dose-verification method to IMRT and IMRS, in which each treatment case was delivered using the sliding-window technique. A pinpoint ion chamber (0.015 cm$^3$; PTW 31006, Germany) was used for absolute point dosimetry, and both absolute point dosimetry and 2D film dosimetry were performed throughout the IMRT and IMRS fields. In the case of IMRT, the gantry and couch angles were fixed to zero. Measurement with a static gantry reduces the QA implementation time but obviously cannot take gantry and couch sagging into account. Since the simplified QA procedure does not represent the actual situation, this procedure certainly validates the consistency of the delivered dose distribution. In the case of IMRS, the actual gantry and couch angles determined by the planning procedure were used to demonstrate the actual situation of the delivery process.

We first consider the IMRT QA results without applying weighting, since they show the effect of optimization more intuitively. Figure 5 shows a listing of an intermediate result of the optimization process. We already mentioned that the dose-difference map is bowl shaped and concave up. Accordingly, the position $X=-1$ mm, $Y=1$ mm is probably the global minimum of this map (at a resolution of 1 mm), indicating that the film has been shifted 1 mm in the $X$ direction and $-1$ mm in the $Y$ direction during the measurement procedure. Here $X$ is defined along the lateral direction and increases to the right, and $Y$ is defined along the axial direction and increases in the gantry direction. The discrete average values in Fig. 3 show that there is the possibility of determining a more accurate global minimum near the already optimized position $X=-1$ mm, $Y=1$ mm. A more precise meaning of the translation is that the global minimum is located between $-2$ and $0$ mm in the $X$ direction and between $0$ and $2$ mm in the $Y$ direction. Therefore the dose distribution with submillimeter resolution can provide us a more accurate global minimum up to the given resolution. If we take into account the submillimeter resolution (particularly one of 0.1 mm), then it is easy to see that there exist several hundred optimal-translation candidates for finding the minimum value. Thus additional iterations are necessary for the optimization procedure, and handling the more detailed dose distributions is more important. Correspondingly, the required computation time increases several-fold, which forces us to consider a different optimization-algorithm strategy. Since the resolution of the film does not affect the accuracy of the optimization method, the number of verification points is of the same order even when we employ submillimeter resolution. As we have already mentioned, the resolution of the calculated dose distribution is directly proportional to the accuracy of the optimization, and that of the measured distribution is proportional to the statistical reliability of dose verification. Our results show that the resolutions used are sufficiently accurate and practically useful in a clinical situation.

| $Y$(mm) \ $X$(mm) | -4.0 | -3.0 | -2.0 | -1.0 | 0.0 | 1.0 | 2.0 | 3.0 | 4.0 |
|---|---|---|---|---|---|---|---|---|---|
| 4.0 | 5.89 | 4.83 | 4.05 | 3.67 | 4.11 | 4.96 | 6.04 | 7.23 | 8.45 |
| 3.0 | 5.28 | 4.10 | 3.15 | 2.71 | 3.24 | 4.23 | 5.42 | 6.68 | 7.94 |
| 2.0 | 4.76 | 3.49 | 2.36 | 1.81 | 2.46 | 3.60 | 4.88 | 6.21 | 7.52 |
| 1.0 | 4.42 | 3.07 | 1.81 | C 1.15 | 1.89 | 3.16 | 4.51 | 5.89 | 7.25 |
| 0.0 | 4.55 | 3.20 | 1.98 | 1.42 | B 2.08 | 3.32 | 4.67 | 6.05 | 7.40 |
| -1.0 | 5.00 | 3.70 | 2.62 | 2.18 | 2.70 | A 3.82 | 5.12 | 6.48 | 7.82 |
| -2.0 | 5.51 | 4.30 | 3.39 | 3.05 | 3.44 | 4.40 | 5.62 | 6.95 | 8.27 |
| -3.0 | 6.11 | 5.02 | 4.24 | 3.97 | 4.28 | 5.10 | 6.20 | 7.47 | 8.77 |
| -4.0 | 6.75 | 5.78 | 5.08 | 4.86 | 5.13 | 5.84 | 6.84 | 8.01 | 9.26 |

Fig. 3 The average absolute dose differences for translations. Indices represent 2D space for the optimization procedure, and the intervals reflect the resolution of 1 mm. The value at position $X=0$, $Y=0$ (point **B**) is the average of absolute dose difference without optimization (or without translation), and other values are the averages of corresponding translations. The minimum average dose difference occurs at $X=-1$, $Y=1$, where the film has translated to the lower-right quadrant of the 2D surface. The dose differences for positions **A**, **B**, and **C** are described in Figs. 6 and 7.

Figs. 4 and 5 show the effect of a tiny shift of the film. In these figures the symbol 'o' denotes that the dose difference (*DD*) is less than 3%, '3' denotes $3\% \leq DD < 5\%$, '5' denotes $5\% \leq DD < 7\%$, and '7' denotes that *DD* is equal or greater than 7% in the corresponding point. A red number indicates that the dose of the calculated distribution is larger than that of the measured one, and a blue one indicates the opposite case.

In the case of Fig. 4(a) (point **B** in Fig. 3) the average value of the absolute dose difference ($\widetilde{\Delta}$) is 2.04%, the absolute value of the maximum dose difference ($DD_{max}$) is 12.38%, and 154 out of 742 total verification points (20.75%) have dose differences over the 3% criterion. For Fig. 4(b) (point **A** in Fig. 3) these values are 3.84%, 22.01%, and 333 points (44.88%), respectively. The optimized verification result is reported on the Fig. 5(a). In Fig. 5(a) $\widetilde{\Delta}$ is 1.24%, $DD_{max}$ is 6.63%, and 40 out of 742 total verification points (5.39%) have dose differences over the 3% criterion. The series of translations demonstrate that a translation even within the LINAC tolerance can generate serious differences in the high-dose-gradient regions, confirming our earlier discussion on the potential problems in these regions.

Fig. 6 shows the weighted result. The average of the weighted absolute dose difference $\widetilde{\Delta}_w$ is 1.03%, $DD_{max}$ is 4.52%, and 23 out of 742 total selected points (3.10%) have differences over the 3% criterion. Comparison of Fig. 5(a) and Fig. 6 reveals that combining weighting and optimization improves the result in the low-dose region and provides a more accurate and meaningful result from an absolute dosimetric viewpoint. Several defects still appear at the high-dose-gradient region despite the use of an optimization method, suggesting that introducing submillimeter resolution would allow more accurate determination of the translation.

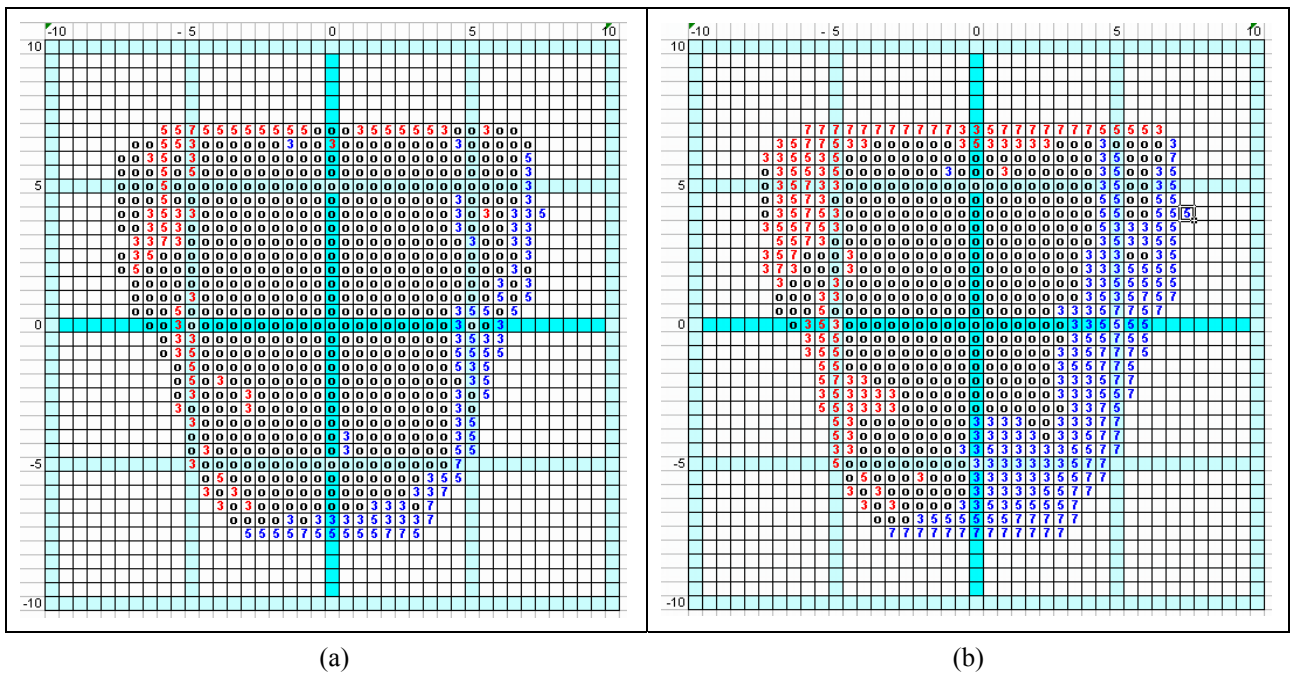

Fig. 4. The point-dose-difference distribution without translation (a; point **B** in Fig. 5) and with a translation of $X=1$ mm, $Y=-1$mm (b; point **A** in Fig. 5). An LC of 30% and a criterion of 3% were used.

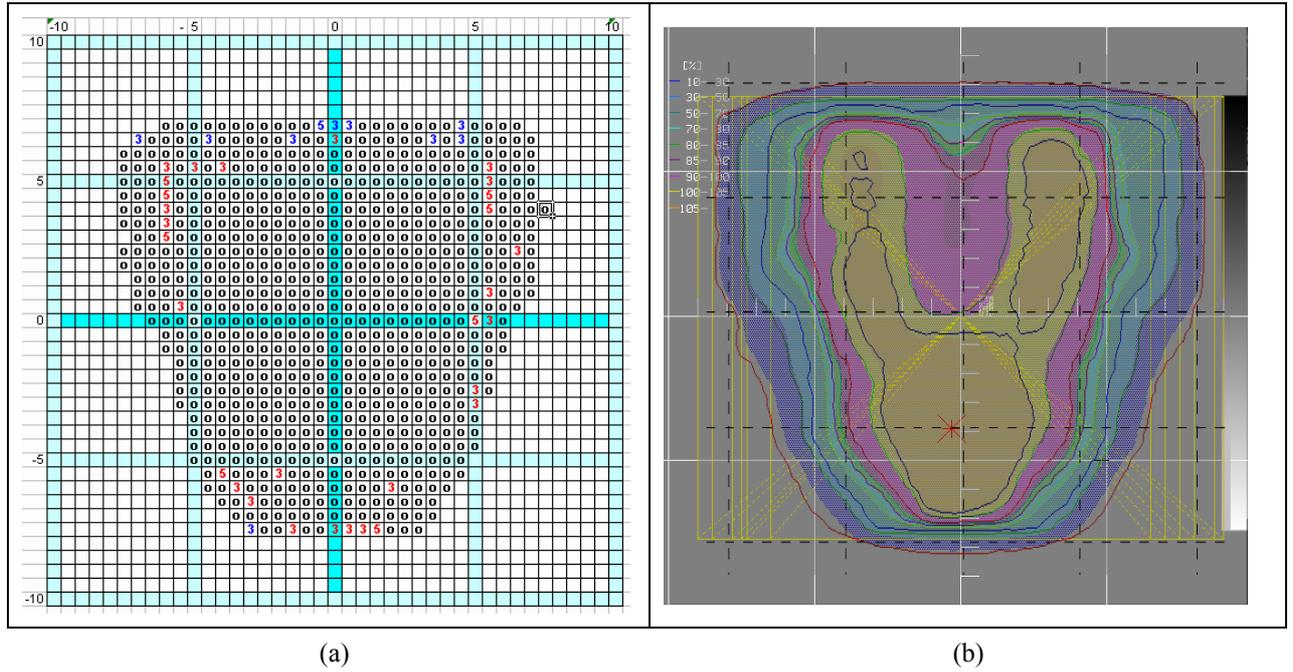

Fig. 5. The point-dose-difference distribution at the optimized positions $X=-1$ mm, $Y=1$ mm (a) and superimposed dose distributions for the planning system and film output (b). In b, the color surface and white grid lines are the result of the planning system, and the contours and dashed grid are from the film output.

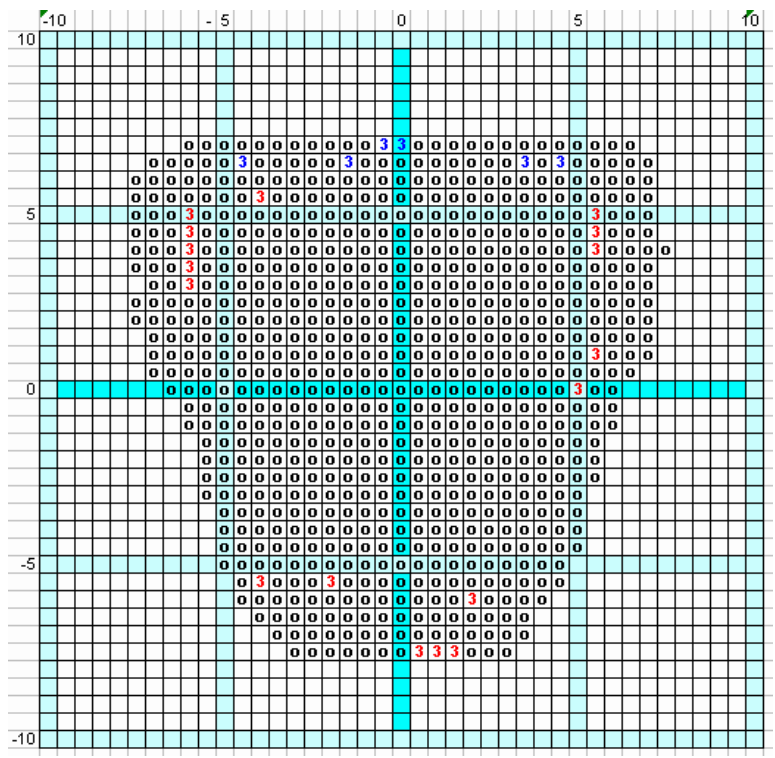

Fig. 6. The distribution maps of the weighted point-dose difference at the optimized position $X=-1$ mm, $Y=-1$ mm. An LC of 30% and a criterion of 3% were used.

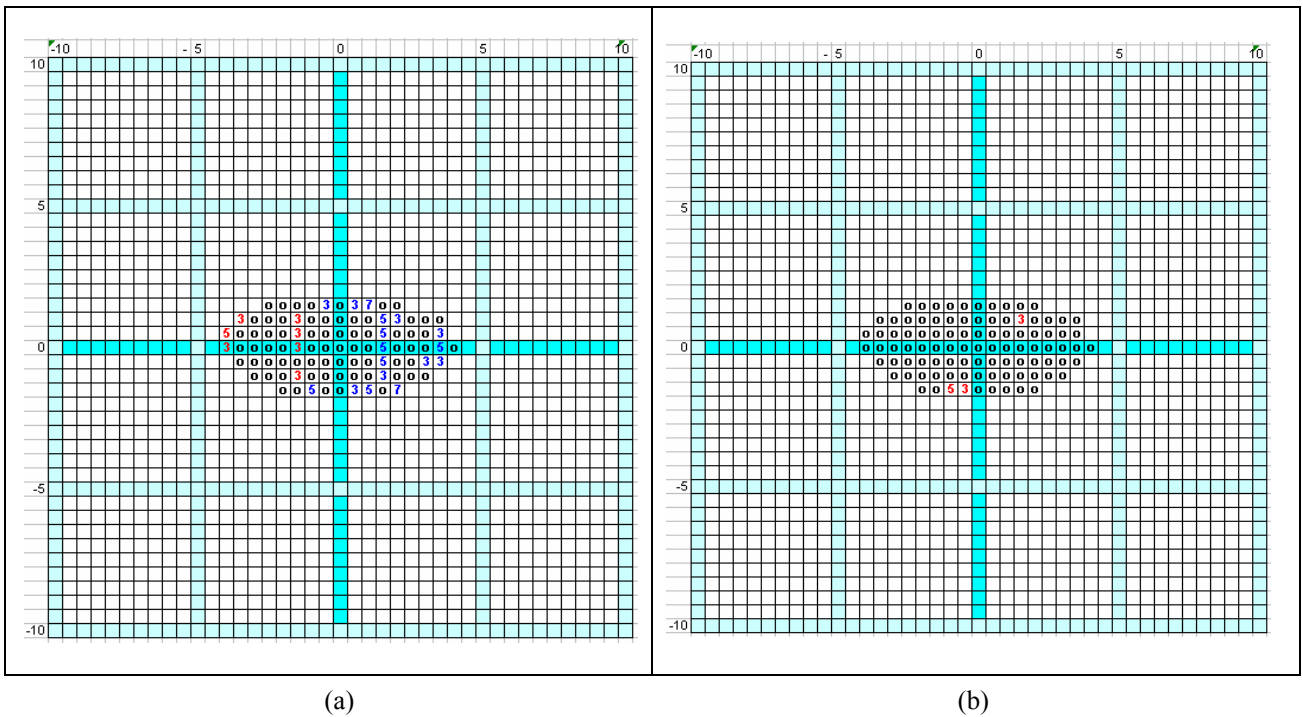

(a)                                                  (b)

Fig. 7. The point-dose-difference distribution without translation (a) and with a translation of $X=0.5$ mm, $Y=0.0$ mm (b). An LC of 30% and a criterion of 3% were used.

    For IMRS, we used calculated and measured dose distributions with resolutions of 0.5 and 5

mm, respectively. Fig. 7(a) shows the result of the IMRS case without optimization and weighting. In this case $\widetilde{\Delta}_w$ is 1.94%, $DD_{max}$ is 7.19%, and 24 out of 95 total verification points (25.3%) have differences over the 3% criterion. Fig. 7(b) shows the result of the IMRS case with optimization and weighting. In this case $\widetilde{\Delta}_w$ is 0.96%, $DD_{max}$ is 6.13%, and 3 out of 95 total verification points (3.16%) have differences over the 3% criterion.

## 4. DISCUSSION AND CONCLUSIONS

In this paper we have described a quantitative dose-verification algorithm that uses a global optimization method. This method of verifying alternative dose maps could also be applied to other types of treatment and dosimeters. There remain many mechanisms by which systematic errors could be generated through the QA procedure. When using film, the film calibration for OD and distortion contribute to the uncertainty of QA procedures. In addition, errors in the mechanical setup such as tiny rotations of the measurement devices, small tilting of the couch, and unavoidable sagging of the gantry also have an effect on the accuracy of the results. Though the errors that we have considered should not change the final result significantly, the analyses help in understanding the relationship between accurate dose verification and systematic errors.

The variation in the OD curves with respect to the continuous calibration changes the film dosimetry result by at most ±7% [18], but the method we developed is not especially sensitive to the absolute density values of the OD curve. Rescaling the measured dose distribution and changing the gradient of the OD curve has the same effect from a relative dosimetric viewpoint. If the OD curves are sufficiently linear (or sufficiently linear parts of OD curves are used) and the OD curves for each calibration are linearly proportional to each other, then we can apply the aforementioned rescaling directly. In the case of EDR2 film the OD curves were sufficiently linear, so that rescaling the measured dose distribution may compensate most of the uncertainty. Nevertheless, frequent and precise OD calibrations are always recommended. The recent development of the daily film calibration method for OD using MLC sequencing files allows an accurate OD curve to be obtained rapidly [18]. Another rigorous method of performing absolute film dosimetry and calibration is described in [19].

We used several parameters to assess our results: $\widetilde{\Delta}_w$, $DD_{max}$, $N$, the number of points with dose differences over 3% in IMRT (5% in IMRS), LC, and distance of optimized translation for each direction. Our results suggest the following guidelines as acceptable criteria for examining the acceptability: $\widetilde{\Delta}_w$ should be smaller than 2% for IMRT (3% for IMRS), the ratio between the number of points with dose differences over 3% in IMRT (5% for IMRS) and $N$ should be smaller than 10%, $DD_{max}$ should be smaller than 10%, and the optimized translation should lie within the tolerance of the LINAC. In addition, a fixed LC (e.g., 30%) provided a consistent analysis of QA results. However, there is no unique choice of acceptable criteria, since these will be influenced by the treatment site and the opinion of the physician.

The last issue to be considered is 3D dose verification of IMRT, which is considered the ultimate goal of IMRT QA. It is obvious that 3D dose verification requires more computing power and a more complicated measurement process. Therefore, reducing the implementation time by the use of an effective optimization algorithm is key to the success of the procedure. As in the 2D case, the development of a user-friendly reporting form for the 3D case is also a key factor. The precise 3D dose-verification procedure based on optimization and film dosimetry is currently under investigation, and will be reported elsewhere [20].

## ACKNOWLEDGEMENTS

This investigation was supported by a research grant from the National Cancer Center, Korea (no. 0110210).

a) Electronic mail: donghyun@ncc.re.kr
b) Corrsponding author: Sung-Yong Park, Ph.D.; electronic mail: cool_park@ncc.re.kr